\documentclass[12pt]{article}

\usepackage{bbm}

\newcommand{\be}{\begin{equation}}
\newcommand{\ee}{\end{equation}}
\newcommand{\ba}{\begin{eqnarray}}
\newcommand{\ea}{\end{eqnarray}}
\newcommand{\no}{\nonumber\\}

\begin{document}

\title{\normalsize \hfill UCRHEP-T386 \\[8mm]
\LARGE
Two Predictive Supersymmetric $S_3 \times Z_2$ Models \\
for the Quark Mass Matrices}
\author{
Lu\'\i s Lavoura\thanks{E-mail: balio@cftp.ist.utl.pt} \\
\small Universidade T\'ecnica de Lisboa
and Centro de F\'\i sica Te\'orica de Part\'\i culas \\
\small Instituto Superior T\'ecnico, 1049-001 Lisboa, Portugal
\\*[4.6mm]
\setcounter{footnote}{6}
Ernest Ma\thanks{E-mail: ma@physics.ucr.edu} \\
\small Physics Department, University of California \\
\small Riverside, California 92521, U.S.A. \\*[3.6mm]
}

\date{February 19, 2005} 

\maketitle

\begin{abstract}
We propose two simple models for the quark mass matrices
which may be implemented through an $S_3 \times Z_2$ symmetry
in a supersymmetric context.
Each model has eight parameters and,
therefore,
makes two independent predictions for the quark mixing matrix.
The first model predicts
$\left| V_{ub} / V_{cb} \right| \approx \sqrt{m_u / m_c} \sim 0.06$
and
$\left| V_{td} / V_{ts} \right| \approx \sqrt{m_d / m_s} \sim 0.23$.
The second model,
in which the forms of the up-type-quark
and down-type-quark mass matrices
are interchanged relative to the first one,
predicts
$\left| V_{ub} / V_{cb} \right| \sim 0.11$
and
$\left| V_{td} / V_{ts} \right| \sim 0.33$.
Both models have $\sin{2 \beta} \sim 0.5$.
\end{abstract}

\newpage

\baselineskip 18pt

\section{Introduction}

The history of models for the quark mass matrices
can be traced back to Weinberg's
original observations~\cite{weinberg} that
\be
\sin{\theta_C} \approx \sqrt {\frac{m_d}{m_s}}
\label{weinberg1}
\ee
(where $\theta_C$ is the Cabibbo angle and $m_d$,
$m_s$ are the down-quark and strange-quark masses,
respectively)
and that mass matrices of the form
\be
M_u = \left( \begin{array}{cc} m_u & 0 \\ 0 & m_c \end{array} \right),
\quad
M_d = \left( \begin{array}{cc} 0 & p \\ p & q \end{array} \right)
\label{weinberg2}
\ee
(where $M_u$ and $M_d$ are the charge $2/3$
and charge $-1/3$ quark mass matrices,
respectively,
while $m_u$ and $m_c$ are the up-quark and charm-quark masses,
respectively)
can account for the aproximate equality~(\ref{weinberg1}).
Nowadays we know that there are three families of quarks
and therefore the matrices~(\ref{weinberg2})
must be generalized in some way.
Three possible generalizations are
\be
M_1 = \left( \begin{array}{ccc}
m_1 & 0 & 0 \\ 0 & m_2 & 0 \\ 0 & 0 & m_3
\end{array} \right),
\quad
M_2 = \left( \begin{array}{ccc}
0 & p & 0 \\ p & q & 0 \\ 0 & 0 & m
\end{array} \right),
\quad
M_3 = \left( \begin{array}{ccc}
0 & x & 0 \\ x & 0 & z \\ 0 & y & w
\end{array} \right).
\label{massmatrices}
\ee
Note that $y \neq z$ is an important condition on $M_3$.
If both $M_u$ and $M_d$ are of the form $M_1$,
then there is no mixing.
If both $M_u$ and $M_d$ are of the form $M_2$,
then there is at least one zero matrix element
in the quark mixing (CKM) matrix $V$.
The possibility that both $M_u$ and $M_d$ are of the form $M_3$
has been recently advocated~\cite{babu}
and shown to follow from a particular realization
of a symmetry group $Q_6$;
however,
it leads to a model with ten---or nine
at best~\cite{babu}---parameters,
which does not have much predictive power
since there are six quark masses
and four parameters of the CKM matrix.
If one of the quark mass matrices is of the form $M_1$
and the other one is of the form $M_2$,
then only two generations mix.
The possibility that $M_u$ is of the form $M_1$
while $M_d$ is (to a good approximation) of the form $M_3$
has been studied by one of us~\cite{ma} a long time ago;
however,
as was shown by the other one os us~\cite{lavoura},
it predicts a rather low value of $m_s/m_d$,
which is disfavored by the data.

In this paper we want to show that
it is quite reasonable to assume one of the quark mass matrices
to be of the form $M_2$ while the other one is of the form $M_3$.
In section~2  we show that this assumption
can be justified by a \emph{supersymmetric} model
with six Higgs doublets (instead of only two)
and a $S_3 \times Z_2$ symmetry.
In section~3 we study in detail the case in which $M_u$
is of the form $M_3$ while $M_d$ is of the form $M_2$,
showing that it predicts
$\left| V_{ub} / V_{cb} \right| \approx \sqrt{m_u / m_c}$
and $\left| V_{td} / V_{ts} \right| \approx \sqrt{m_d / m_s}$.
In section~4 we study the opposite case,
where $M_u$ is of the form $M_2$ and $M_d$ is of the form $M_3$,
emphasizing that it leads to much higher values of both
$\left| V_{ub} / V_{cb} \right|$
and $\left| V_{td} / V_{ts} \right|$.
We summarize our findings in section~5.

\section{Justification for the form of the mass matrices}
\label{justification}

Consider a supersymmetric model with six Higgs doublets,
three of them ($H^i_{1,2,3}$) with weak hypercharge $+1/2$
and the other three ($H^i_{4,5,6}$) with hypercharge $-1/2$
($i$ is a gauge $SU(2)$ index).
The three quark doublets,
carrying weak hypercharge $1/6$,
are $Q^i_{1,2,3}$;
the quark singlets with hypercharge $-2/3$
are denoted $p_{1,2,3}$
and the singlets with hypercharge $1/3$
are denoted $n_{1,2,3}$.
Let there be the following two symmetries $A$ and $B$ in the model:
\be
A: \quad \begin{array}{lll}
\left\{ \begin{array}{rcl}
H_1 &\to& \omega H_1, \\ H_2 &\to& \omega^2 H_2,
\end{array} \right.
&
\left\{ \begin{array}{rcl}
H_4 &\to& \omega H_4, \\ H_5 &\to& \omega^2 H_5,
\end{array} \right.
&
\\
\left\{ \begin{array}{rcl}
Q_1 &\to& \omega Q_1, \\ Q_2 &\to& \omega^2 Q_2,
\end{array} \right.
&
\left\{ \begin{array}{rcl}
p_1 &\to& \omega p_1, \\ p_2 &\to& \omega^2 p_2,
\end{array} \right.
&
\left\{ \begin{array}{rcl}
n_1 &\to& \omega n_1, \\ n_2 &\to& \omega^2 n_2,
\end{array} \right.
\end{array}
\ee
where $\omega \equiv \exp{\left( 2 i \pi / 3 \right)}$;
and
\be
B: \quad \begin{array}{ccc}
H_1 \leftrightarrow H_2,
&
H_4 \leftrightarrow H_5,
&
\\
Q_1 \leftrightarrow Q_2,
&
p_1 \leftrightarrow p_2,
&
n_1 \leftrightarrow n_2.
\end{array}
\ee
Obviously,
$A$ and $B$ do not commute,
and together they generate a representation of the group $S_3$.
Besides this $S_3$,
we impose a reflection ($Z_2$) symmetry $C$ under which
\be
C: \quad H_{1,2} \to - H_{1,2},
\quad Q_3 \to -Q_3, \quad p_3 \to - p_3, \quad n_3 \to - n_3.
\ee
The full horizontal symmetry group is thus $S_3 \times Z_2$.
The part of the superpotential yielding the usual Yukawa couplings is
\ba
\!\!\!\! & & \epsilon^{ij} \left[
f_1 \left( Q_1^i p_2 + Q_2^i p_1 \right) H_3^j
+
f_2 \left( Q_1^i H_2^j + Q_2^i H_1^j \right) p_3
+
f_3 Q_3^i \left( p_1 H_2^j + p_2 H_1^j \right)
+
f_4 Q_3^i p_3 H_3^j 
\right.
\no
\!\!\!\! & & \left.
+
f_5 H_6^i \left( Q_1^j n_2 + Q_2^j n_1 \right)
+
f_6 \left( H_4^i Q_1^j n_1 + H_5^i Q_2^j n_2 \right)
+
f_7 H_6^i Q_3^j n_3 
\right],
\label{W}
\ea
where $\epsilon^{ij}$ is an antisymmetric $2 \times 2$ matrix
responsible for contracting the doublet $SU(2)$ indices
into an $SU(2)$ singlet,
and $f_{1,\ldots,7}$ are dimensionless coupling constants.
After the doublets $H_{1,\ldots,6}$
acquire vacuum expectation values (VEVs) $v_{1,\ldots,6}$,
respectively,
the quark mass matrices are
\be
M_u = \left( \begin{array}{ccc}
0 & f_1 v_3 & f_2 v_2 \\
f_1 v_3 & 0 & f_2 v_1 \\
f_3 v_2 & f_3 v_1 & f_4 v_3
\end{array} \right),
\quad
M_d = \left( \begin{array}{ccc}
f_6 v_4 & f_5 v_6 & 0 \\
f_5 v_6 & f_6 v_5 & 0 \\
0 & 0 & f_7 v_6
\end{array} \right).
\ee
If $v_2$ and $v_4$ vanish but the other four VEVs are non-zero,
$M_u$ is of the form $M_3$ and $M_d$ is of the form $M_2$,
as desired.
It is not possible to enforce this pattern of VEVs
in a non-supersymmetric model,
since the term in the Higgs potential
$\left( H_3^\dagger H_1 \right) \left( H_3^\dagger H_2 \right)$,
which is allowed by the symmetry $S_3 \times Z_2$,
forces $v_2$ to become non-zero
once $v_1$ and $v_3$ are simultaneously non-zero;
similarly,
$\left( H_6^\dagger H_4 \right) \left( H_6^\dagger H_5 \right)$
generates a non-zero $v_4$ out of non-vanishing $v_5$ and $v_6$.
However,
in a supersymmetrized version of the model
such terms are absent from the tree-level Higgs potential.
Indeed,
the supersymmetric scalar potential is composed of three parts:
$D$-terms,
$F$-terms,
and soft-breaking terms.
The $D$-terms only amount,
in our case,
to
\be
\left\langle 0 \left| V_D \right| 0 \right\rangle =
\frac{g^2 + {g^\prime}^2}{8} \left(
\left| v_1 \right|^2 + \left| v_2 \right|^2 + \left| v_3 \right|^2
- \left| v_4 \right|^2 - \left| v_5 \right|^2 - \left| v_6 \right|^2
\right)^2,
\ee
where $V_D$ is the $D$-terms scalar potential
and $g$, $g^\prime$ are the $SU(2)$ and $U(1)$
gauge coupling constants,
respectively.
The $F$-terms scalar potential,
$V_F$,
is determined by the superpotential $W$,
which is the sum of the trilinear terms of~(\ref{W})
with the unique bilinear term $\mu H_3 H_6$
allowed by the horizontal symmetry $S_3 \times Z_2$;
one has $V_F = \sum_a F_a F_a^\ast$,
where $F_a \equiv \partial W / \partial A_a$
is the partial derivative of the superpotential
relative to any superfield which occurs in it.
It is easy to see that there are no terms in $V_F$
containing four Higgs scalar fields.
Finally,
the soft-breaking terms in the scalar potential,
which by definition are never quartic,
can be assumed to respect a symmetry
$H_2 \to - H_2,\ H_4 \to - H_4$,
so that they will never generate non-zero VEVs for $H_2$ and $H_4$.

The terms
$\left( H_3^\dagger H_1 \right) \left( H_3^\dagger H_2 \right)$
and 
$\left( H_6^\dagger H_4 \right) \left( H_6^\dagger H_5 \right)$
can be generated at one-loop level.
In the exact supersymmetric limit, 
the bosonic and fermionic contributions to the loops
cancel exactly;
but the cancellation is not perfect when supersymmetry is broken.
Therefore, 
those terms will be present but naturally suppressed,
so that one may assume that $v_2 << v_1$
and $v_4 << v_5$ to an excellent approximation.

\section{First model}
\label{first}

Let the quark mass matrices be
\be
M_u = \left( \begin{array}{ccc}
0 & x & 0 \\ x & 0 & z \\ 0 & y & w
\end{array} \right),
\quad
M_d = \left( \begin{array}{ccc}
0 & p & 0 \\ p & q & 0 \\ 0 & 0 & m_b e^{i \theta}
\end{array} \right),
\label{M1}
\ee
where $m_b$ is the bottom-quark mass
and $\theta$ is a physically meaningless phase.
Defining the Hermitian matrices
\ba
{\cal H}_u &\equiv& M_u M_u^\dagger, \label{Hu}
\\
{\cal H}_d &\equiv& M_d M_d^\dagger, \label{Hd}
\ea
their diagonalization proceeds through unitary matrices
$V_u$ and $V_d$ as
\ba
V_u^\dagger {\cal H}_u V_u &=&
\mbox{diag} \left( m_u^2,\, m_c^2,\, m_t^2 \right),
\label{diagu}
\\
V_d^\dagger {\cal H}_d V_d &=&
\mbox{diag} \left( m_d^2,\, m_s^2,\, m_b^2 \right),
\label{diagd}
\ea
and the CKM matrix is
\be
V = V_u^\dagger V_d.
\label{V}
\ee
In the case of the matrix $M_d$ of~(\ref{M1}),
it is clear that
\be
V_d = \left( \begin{array}{ccc}
\cos{\alpha_d} & \sin{\alpha_d} & 0 \\
- e^{i \psi} \sin{\alpha_d} & e^{i \psi} \cos{\alpha_d} & 0 \\
0 & 0 & 1
\end{array} \right),
\label{Vd}
\ee
where
\be
\sin{\alpha_d} = \sqrt{\frac{m_d}{m_s + m_d}},
\quad
\cos{\alpha_d} = \sqrt{\frac{m_s}{m_s + m_d}}.
\label{alphad}
\ee
The phase $\psi$ is chosen to be
$\psi = \arg{\left( x y^\ast z^\ast w p^\ast q \right)}$,
and then the matrix $M_u$ can be taken to be real
with non-negative matrix elements.
The unitary matrix $V_u$ is real.

It follows from~(\ref{V}) and~(\ref{Vd})
that the third column of $V$ is identical with
the third column of $V_u^\dagger$.
Let us denote
$\left| V_{ub} \right|^2 \equiv \alpha$,
$\left| V_{cb} \right|^2 \equiv \beta$,
and $\left| V_{tb} \right|^2 \equiv \gamma$;
obviously,
$\alpha + \beta + \gamma = 1$.
One derives from~(\ref{diagu}) that
\ba
a \ \equiv \
m_u^2 \alpha + m_c^2 \beta + m_t^2 \gamma &=&
\left| y \right|^2 + \left| w \right|^2,
\label{aa} \\
c \ \equiv \
m_u^4 \alpha + m_c^4 \beta + m_t^4 \gamma
- \left( m_u^2 \alpha + m_c^2 \beta + m_t^2 \gamma \right)^2 &=&
\left| x y \right|^2 + \left| z w \right|^2.
\label{cc}
\ea
We may put these equations together with
\ba
b \ \equiv \
m_u^2 \left( 1 - \alpha \right)
+ m_c^2 \left( 1 - \beta \right)
+ m_t^2 \left( 1 - \gamma \right)
&=& 2 \left| x \right|^2 + \left| z \right|^2,
\label{bb} \\
d \ \equiv \
m_u^2 m_c^2 + m_u^2 m_t^2 + m_c^2 m_t^2 &=&
\left| x \right|^4
+ \left| x y \right|^2
\no & &
+ \left| x z \right|^2
+ 2 \left| x w \right|^2
+ \left| y z \right|^2
\label{dd}
\ea
and obtain a system~(\ref{aa})--(\ref{dd})
which has the exact solution
\ba
\left| z \right|^2 &=&
\sqrt{b^2 + 4 \left( a b - c - d \right)},
\label{z} \\
\left| x \right|^2 &=&
\frac{b - \left| z \right|^2}{2},
\label{x} \\
\left| y \right|^2 &=&
\frac{2 \left( a \left| z \right|^2 - c \right)}
{3 \left| z \right|^2 - b},
\label{y} \\
\left| w \right|^2 &=&
\frac{a \left| z \right|^2 + 2 c - a b}
{3 \left| z \right|^2 - b}.
\label{w}
\ea
If one furthermore uses
\be
m_u^2 m_c^2 m_t^2 = \left| x^2 w \right|^2,
\label{determinant}
\ee
then one ends up with one equation
connecting the masses of the charge $2/3$ quarks to $\alpha$,
$\beta$,
and $\gamma$.
That equation may be solved numerically.
One can in this way derive,
in an exact numerical fashion,
the consequences of the mass matices~(\ref{M1}).

Our model~(\ref{M1}) for the quark mass matrices
is an eight-parameter model---the parameters
are the moduli of $x$,
$y$,
$z$,
$w$,
$p$,
and $q$,
together with $m_b$ and the phase $\psi$.
It should therefore produce two independent predictions.
One may perform an approximate diagonalization
of the matrix ${\cal H}_u$
by using the following formulae,
originally given in~\cite{ma}:
\ba
m_t &\approx& \sqrt{y^2 + w^2},
\label{mt} \\
m_c &\approx& \frac{y z}{\sqrt{y^2 + w^2}},
\label{mc} \\
m_u &\approx& \frac{x^2 w}{y z},
\label{mu} \\
\left( V_u \right)_{23} &\approx& \frac{z w}{y^2 + w^2},
\label{vu23} \\
\left( V_u \right)_{12} &\approx& - \frac{x w}{y z},
\label{vu12} \\
\left( V_u \right)_{13} &\approx& \frac{x y}{y^2 + w^2}.
\label{vu13}
\ea
From the orthogonality of $V_u$
one finds that
\ba
\left( V_u \right)_{32} &\approx& - \left( V_u \right)_{23},
\label{vu32} \\
\left( V_u \right)_{21} &\approx& - \left( V_u \right)_{12},
\label{vu21} \\
\left( V_u \right)_{31} &\approx& - \left( V_u \right)_{13}
+ \left( V_u \right)_{12} \left( V_u \right)_{23}.
\label{vu31}
\ea
We note that $y^2 \ll w^2$ is necessary in order to fit the data.
We then obtain
\be
\frac{V_{ub}}{V_{cb}}
= \frac{\left( V_u \right)_{31}}{\left( V_u \right)_{32}}
\approx \frac{x w}{y z} \approx \sqrt{\frac{m_u}{m_c}}.
\ee
This may be considered as the first prediction of the model.
One also finds that
\be
\frac{\left( V_u \right)_{13}}{\left( V_u \right)_{23}}
\approx \frac{x y}{z w} \approx
\frac{m_c^2}{m_t^2} \sqrt{\frac{m_u}{m_c}}
\frac{1}{\left| V_{cb} \right|^2},
\ee
which is very small,
of order $10^{-3}$.
It follows from~(\ref{V})--(\ref{alphad}),
when $\left( V_u \right)_{13}$ is neglected,
that
\be
\frac{V_{td}}{V_{ts}} \approx - \tan{\alpha_d}
= - \sqrt{\frac{m_d}{m_s}}.
\ee
This may be considered as a second prediction of the model.

The predictions
\ba
\left| V_{ub} \right|^2 &\approx&
\left| V_{cb} \right|^2 \frac{m_u}{m_c},
\label{uc}
\\
\left| V_{td} \right|^2 &\approx&
\left| V_{ts} \right|^2 \frac{m_d}{m_s}
\label{ds}
\ea
are not new
in the history of mass-matrix models and \textit{Ans\"atze}:
they were first arrived at several years ago~\cite{branco}
in the context of a very peculiar \textit{Ansatz}
for the quark mass matrices.
It should be stressed that that \textit{Ansatz}
was,
and remains,
completely unjustified in terms of a full model.
This is not what happens with the matrices~(\ref{M1}).
Numerically,
one finds that the approximate forms~(\ref{uc}) and~(\ref{ds})
for the predictions of the present model are quite good:
the difference between the left- and right-hand sides
of~(\ref{uc}) is about 5\% of each of them,
and the approximation~(\ref{ds}) is much better,
holding at the 0.05\% level.

The prediction~(\ref{uc}) yields a rather small value for
$\left| V_{ub} / V_{cb} \right| \sim 0.06$,
which is in rough agreement with the data
on exclusive semileptonic decays of the bottom quark,
but not with the data on inclusive decays \cite{UTfit}.
Such a low value for $\left| V_{ub} / V_{cb} \right|$
is \emph{incompatible}
with the measured value of $\sin{2 \tilde \beta}$,
if we interpret that measurement
in the framework of the standard model,
where $\tilde \beta = \beta \equiv \arg{\left( - V_{cd} V_{tb}
V_{cb}^\ast V_{td}^\ast \right)}$.
Indeed,
in our model we obtain $\sin{2 \beta} = 0.52$
for the central values of the quark masses~\cite{gasser},
or $\sin{2 \beta} = 0.60$ if the light-quark masses $m_u$,
$m_d$,
and $m_s$ are taken at their upper bounds~\cite{gasser};
experimentally~\cite{browder},
on the other hand,
$\sin{2 \tilde \beta} = 0.736 \pm 0.049$.
However,
in the present model there are extra contributions
to $B_d^0$--$\bar B_d^0$ mixing,
for instance from box diagrams with intermediate charged scalars,
through the terms with coefficients $f_2$ and $f_3$
in~(\ref{W}).\footnote{There are also extra contributions
to $K^0$--$\bar K^0$ mixing,
which affect indirectly the experimental process
$B^0_d$/$\bar B^0_d \to J/\psi K_S$
through which $\sin{2 \tilde \beta}$ is measured.
Those contributions are not necessarily small,
since the Yukawa couplings may be large
if the VEVs of the involved Higgs doublets are sufficiently small.}

From the moduli of the CKM matrix elements
one may compute the Wolfenstein~\cite{wolfenstein} parameters
$\rho$ and $\eta$ by using
\ba
\left| V_{ub} \right|^2 &=& \left| V_{us} V_{cb} \right|^2
\left( \rho^2 + \eta^2 \right),
\\
\left| V_{td} \right|^2 - \left| V_{ub} \right|^2 &\approx&
\left| V_{us} V_{cb} \right|^2
\left( 1 - 2 \rho \right).
\ea
In the present model one obtains $\rho \sim 0.06$
and $\eta \sim 0.27$.
If the light-quark masses are taken at their upper bounds,
one can reach $\rho$ as high as $0.08$
and $\eta$ as high as $0.31$;
in the best fits of the standard model~\cite{UTfit}
one usually obtains higher values for $\rho$ and $\eta$,
but once again we should remember that our model is \emph{not}
the standard model.

\section{Second model}
\label{second}

Instead of~(\ref{M1}),
let the quark mass matrices be
\be
M_d = \left( \begin{array}{ccc}
0 & x & 0 \\ x & 0 & z \\ 0 & y & w
\end{array} \right),
\quad
M_u = \left( \begin{array}{ccc}
0 & p & 0 \\ p & q & 0 \\ 0 & 0 & m_t e^{i \theta}
\end{array} \right),
\label{M2}
\ee
where $m_t$ is the top-quark mass and $\theta$ is,
as before,
a physically meaningless phase.
Equations~(\ref{Hu})--(\ref{V}) remain valid.
The matrix $V_u$ is now given by
\be
V_u^\dagger = \left( \begin{array}{ccc}
\cos{\alpha_u} & - e^{- i \psi} \sin{\alpha_u} & 0 \\
\sin{\alpha_u} & e^{- i \psi} \cos{\alpha_u} & 0 \\
0 & 0 & 1
\end{array} \right),
\label{Vu2}
\ee
where
\be
\sin{\alpha_u} = \sqrt{\frac{m_u}{m_c + m_u}},
\quad
\cos{\alpha_u} = \sqrt{\frac{m_c}{m_c + m_u}}.
\label{alphau}
\ee
The phase $\psi = \arg{\left( x y^\ast z^\ast w p^\ast q \right)}$,
and then the matrix $M_d$ in~(\ref{M2})
can be taken to be real with non-negative matrix elements.
The matrix $V_d$ is real orthogonal.

One may use again the original equations of~\cite{ma}.
Defining $r \equiv y/w$ one has
\ba
m_b &\approx& w \sqrt{1 + r^2},
\label{mb} \\
m_s &\approx& \frac{r z}{\sqrt{1 + r^2}},
\label{ms} \\
m_d &\approx& \frac{x^2}{r z},
\label{md} \\
\left( V_d \right)_{23} \approx - \left( V_d \right)_{32}
&\approx& \frac{z}{w \left( 1 + r^2 \right)},
\label{vd23} \\
\left( V_d \right)_{12} \approx - \left( V_d \right)_{21}
&\approx& - \frac{x}{r z},
\label{vd12} \\
\left( V_d \right)_{13} &\approx&
\frac{r x}{w \left( 1 + r^2 \right)},
\label{vd13} \\
\left( V_d \right)_{31} &\approx&
- \frac{r x}{w \left( 1 + r^2 \right)}
- \frac{x}{r w \left( 1 + r^2 \right)}.
\label{vd31}
\ea
Using~(\ref{V}),
(\ref{Vu2}),
and the smallness of the angle $\alpha_u$,
one then has
\ba
V_{cb} &\approx& e^{- i \psi} \frac{m_s}{r m_b},
\label{vcb} \\
V_{us}
&\approx&
- \frac{1}{\sqrt[4]{1 + r^2}} \sqrt{\frac{m_d}{m_s}}
- e^{- i \psi} \sqrt{\frac{m_u}{m_c}},
\label{vus} \\
\frac{V_{td}}{V_{ts}}
&\approx& \sqrt{1 + r^2}\, \sqrt{\frac{m_d}{m_s}},
\label{vtdvts} \\
\frac{V_{ub}}{V_{cb}}
&\approx&
e^{i \psi} \frac{r^2}{\sqrt[4]{1 + r^2}}\, \sqrt{\frac{m_d}{m_s}}
- \sqrt{\frac{m_u}{m_c}}.
\label{vubvcb}
\ea
This model is close to the original $S_3 \times Z_3$ model
of~\cite{ma} except for the fact that here $M_u$ is not diagonal.
In practice,
this amounts to the $\sqrt{m_u / m_c}$ corrections
in~(\ref{vus}) and~(\ref{vubvcb}).\footnote{In
the $S_3 \times Z_3$ model there were corrections
of order $m_u / m_c$,
hence much smaller,
to the $M_d$ of~(\ref{M2}),
while $M_u$ was kept diagonal.}
Since $r \approx m_s / \left( m_b \left| V_{cb} \right| \right)
\sim 0.8$,
the first contribution to $V_{us}$
in the right-hand side of~(\ref{vus})
is too small by itself alone to fit $\left| V_{us} \right|$ well,
as was pointed out in~\cite{lavoura}.
But the $\sqrt{m_u / m_c}$ term in the right-hand side of~(\ref{vus})
fixes the problem and,
indeed,
one finds $\psi \sim 60^\circ$.
Then,
from~(\ref{vtdvts}) and~(\ref{vubvcb}) one obtains
the predictions $\left| V_{td} / V_{ts} \right| \sim 0.29$
and $\left| V_{ub} / V_{cb} \right| \sim 0.11$,
respectively.
One sees that both ratios are much larger in this model
than in the one of the previous section.

Numerically,
this model for the quark mass matrices yields,
for the central values of the quark masses~\cite{gasser},
$\sin{2 \beta} = 0.46$,
$\rho = - 0.37$,
and $\eta = 0.34$.
It is clear that $\sin{2 \beta}$ is smaller than in the standard model,
also $\rho$ is much smaller than in the standard model's best fit.
By tampering a bit with the quark masses one may easily
render $\eta$ and $\sin{2 \beta}$ about $0.1$ larger,
and $\rho$ about $0.1$ smaller.

\section{Conclusions}
\label{conclusions}

We have shown in this paper that,
by means of a horizontal symmetry $S_3 \times Z_2$,
it is possible to construct models for the quark mass matrices
such that one of those matrices is of the type $M_2$
and the other one is of the type $M_3$ in~(\ref{massmatrices}).
The quartic terms in the scalar potential
must be constrained enough to prevent some Higgs doublets
from acquiring a VEV,
and that can be achieved \emph{only in a supersymmetric context}.

Both the model in which $M_u$ is of the type $M_3$
and $M_d$ is of the type $M_2$,
and the model in which $M_u$ is of the type $M_2$
and $M_d$ is of the type $M_3$,
are able to fit the CKM matrix reasonably well
using \emph{the central values for the quark mass ratios}.
The models feature eight parameters to fit ten experimental quantities,
thus they \emph{are predictive}.
With the central values of the quark mass ratios one obtains
$\left| V_{ub} / V_{cb} \right| \approx 0.06$,
$\left| V_{td} / V_{ts} \right| \approx 0.23$,
$\rho \approx 0.06$,
$\eta \approx 0.27$,
and $\sin{2 \beta} \approx 0.52$ in the first model;
the second model yields
$\left| V_{ub} / V_{cb} \right| \approx 0.11$,
$\left| V_{td} / V_{ts} \right| \approx 0.33$,
$\rho \approx - 0.37$,
$\eta \approx 0.34$,
and $\sin{2 \beta} \approx 0.46$.
By changing the quark mass ratios
one can change these values somewhat,
but not much.
The exact value of the top-quark mass
is mostly irrelevant for the models,
provided it is high enough.
The value of $\sin{2 \beta}$
always comes out lower than the experimental result,
indicating that substantial new-physics contributions
to $B_d^0$--$\bar B^0_d$ mixing
and/or $K^0$--$\bar K^0$ mixing must occur
if these models are realized in Nature.


\paragraph{Acknowledgements:}
The work of L.L.\ was supported by the Portuguese
\textit{Funda\c c\~ao para a Ci\^encia e a Tecnologia}
through the projects POCTI/FNU/37449/2001
and U777--Plurianual.
The work of E.M.\ was supported in part
by the U.S.\ Department of Energy
under Grant No.\ DE-FG03-94ER40837.


\end{document}